# Magneto-dielectric Effect in Relaxor Dipolar Glassy $Tb_2CoMnO_6$ Film

R. Mandal[1,2*], M. Chandra[3], V. Roddatis[4], M. Tripathi[3], R. Rawat[3], R.J. Choudhary[3] and V. Moshnyaga[2*]

[1]*Department of Physics, Indian Institute of Science Education and Research, Pune 411008, India*

[2]*Erstes Physikalisches Institut, Georg-August-Universität Göttingen, Friedrich-Hund-Platz 1, 37077 Göttingen, Germany*

[3]*UGC-DAE Consortium for Scientific Research, Indore Centre, University Campus, Khandwa Road, Indore 452017, India*

[4]*Institut für Materialphysik, Georg-August-Universität Göttingen, Friedrich-Hund-Platz 1, 37077 Göttingen, Germany*

*E-mail; vmosnea@gwdg.de ; rajesh.mandal@students.iiserpune.ac.in

### ABSTRACT

We report magneto-dielectric properties of partially B-site ordered monoclinic $Tb_2CoMnO_6$ double perovskite thin film epitaxially grown by metalorganic aerosol deposition technique. Transmission electron microscopy and electron energy loss spectroscopy mapping shows the presence and distribution of both $Co^{2+}$ and $Co^{3+}$ ions in the film, evidencing a partial B-site disorder, which was further confirmed by the observation of reduced saturation magnetization at low temperatures. The ferromagnetic Curie temperature, $T_C$=110 K, is slightly higher as compared to the bulk value (100 K) probably due to an in plane tensile strain. Remarkably, a short range ordering of spins at $T^* \sim 190$ K$\gg T_C$ was established and assigned to the B-site disorder in the film. Two different dielectric relaxation peaks have been observed; they merge at the same temperature $T^*$ of short range spin correlations. Moreover, an unexpected high temperature dipolar relaxor-glass-like transition at $T \sim T^*$ was observed, at which a coupling to short range magnetic correlations results in a 4% magneto-dielectric coupling.

## I. INTRODUCTION

Magneto-dielectrics and magneto-electric materials with the coupled magnetic and electric dipolar order parameters are of fundamental as well as of technological importance. Rare-earth-based perovskite oxides have been proved to be potential candidates for the next generation memory and spintronic device applications[1–7]. $A_2BB`O_6$ (A is a rare earth cation, B & B` are transition metal ions) double perovskites with the layered $ABO_3/AB'O_3$ cation-ordered structure along the [111] axis represent themselves an emerging and auspicious platform to study strong electronic correlations, complex magnetic structure, spin-lattice interaction and magneto-dielectric coupling[8,9].

$R_2(Co/Ni)MnO_6$ (where R= La to Lu) system having a monoclinic structure with $P2_1/n$ space group is especially attractive as they possess a ferromagnetic insulating (FMI) behavior, allowing a high temperature magneto-dielectric coupling. FMI originates from a 180° super exchange interaction between high spin $Co^{2+}/ Ni^{2+}$ and $Mn^{4+}$ ions according to the second Goodenough-Kanamori-Anderson rule[10–12]. The FM ordering as well as dielectric behaviour depend strongly on the B-site ordering which controls the super exchange interaction along with hoping of charge. The fully and partially B-site ordered $La_2CoMnO_6$ has been well explored due to a reasonably high ferromagnetic Curie temperature, $T_C$=230 K[13], spin-phonon coupling[11] and weak magneto-dielectric effect (3%)[14]. Partially disordered $La_2NiMnO_6$ has been revealed as a promising multi-glass material where two different glassy states (spin and dipolar) are observed simultaneously with a high magneto-dielectric coupling constant ($\epsilon_{MD} = \frac{\epsilon'(8T)-\epsilon'(0T)}{\epsilon'(0T)} \times 100$) ~16% at room temperature[15]. Considering the smaller radii rare earth ions in the A-site (R=Pr to Lu) the ferromagnetic $T_C$ decreases reasonably (48 K for Lu)[8] along with spin-phonon interaction[16]. The trend in the dielectric behavior also remains unchanged from bigger to smaller A-site cations as found in bulk $(La/Tb/Y)_2CoMnO_6$ where dielectric constant decreases monotonically with lowering temperature[17]. The La$_2$CoMnO$_6$ (LCMO) from the concerned rare earth double perovskite family has been reported to possess a large dielectric constant at room temperature, which gradually decreases with lowering the temperature[17]. Taking ions with smaller ionic radii, like Y$^{3+}$ and Tb$^{3+}$ in the A-site, the trend of decreasing dielectric constant monotonically towards lowering temperature remains similar with no significant deviation[18]. The overall behaviour is the same in case of epitaxial LCMO thin films though the dielectric constant becomes very low[14]. A controllable disorder in a perovskite or double perovskite system can create a dipolar glass that can couple with its magnetic subsystem, inducing a novel magneto-dielectric or rare multi-glass behaviour[19]. Tb$^{3+}$ with a small cation radius in the A-site has a special significance as it can tune and stabilize a novel hexagonal structure in a strained thin film, which cannot sustain in the bulk form as observed in TbMnO$_3$[20]. With all these experimental observations and intuitions, the Tb-based double perovskite thin films with controlled B-site (partial) ordering could be suggested as an important and exclusive playground for studying novel magnetic and dielectric transitions along with possible coupling between them.

Here we report the epitaxial growth of monoclinic Tb$_2$CoMnO$_6$/Nb:SrTiO$_3$(100) thin film by using a metalorganic aerosol deposition (MAD) technique[21]. The established partial B-site disorder in the film is accompanied by a short range ordering of spins as manifested by the deviation from a standard Curie-Weiss law at T*~190 K, i.e. far above T$_C$=110 K. Moreover, an unexpected high temperature dipolar relaxor-glass-like transition at T* with a 4% magneto-dielectric coupling was observed.

## II. EXPERIMENTAL SECTION

Tb$_2$CoMnO$_6$ (TCMO) films have been grown by MAD technique on commercial electrically conducting 0.5% Nb-doped SrTiO$_3$(100) (NSTO) substrates (Crystal GmbH). Acetylacetonates of Tb, Mn and Co were used as precursors. Precursor solutions in dimethylformamide (DMF) with concentration 0.02 M (for both Co and Mn-precursor) and molar ratio Tb/Mn=1.1 were prepared. The films with thickness, d=80 nm, were grown by spraying the precursor solution by dried compressed air onto a substrate heated to $T_{sub}$~900°C with an average growth rate of v=15 nm/min and were cooled down to room temperature in 20 min after deposition. Magnetization with respect to temperature and applied magnetic field parallel to the film surface (IP) was measured using commercial 7T-SQUID-VSM (Quantum Design Inc., USA) system. Magnetization vs. temperature was measured following the conventional protocols of zero field cooled warming (ZFC) and field cooled warming (FCW) cycles in the presence of applied magnetic field $\mu_0H$=100 Oe. The local structure of TCMO films was studied by Scanning Transmission Electron Microscopy (STEM) and electron energy loss spectroscopy (EELS) using an FEI Titan 80-300 G2 environmental transmission electron microscope (ETEM), operated at an acceleration voltage of 300 kV. TEM lamellas were prepared by a Focused Ion Beam (FIB) lift-out technique using a Thermo-Fischer (former FEI) Helios 4UC instrument. The temperature- and magnetic-field-dependent complex dielectric measurements were performed using a home-made insert coupled with 9 T superconducting magnet and a Keysight E4980A LCR-meter operating at frequency range f=20 Hz-2MHz.

### III. RESULTS AND DISCUSSION

#### A. Magnetism

Temperature dependence of the ZFC & FC magnetic susceptibility, $\chi(T)$, of the TCMO film, measured for $\mu_0H$=100 Oe, is shown in Fig. 1. One can see a phase transition, at $T_C$=110 K, below which the long range ferromagnetic ordering develops due the super exchange interaction of $Co^{2+}$ and $Mn^{4+}$ ions[12]. The transition seems to be of a second order as it is apparently not sharp and no warming/cooling hysteresis was observed. The transition temperature in TCMO film is a bit higher than that observed in Tb$_2$CoMnO$_6$ single crystal (100 K)[22]. The reason is not very clear, but, since the calculated from X-ray diffraction pattern (see Supplementary Information (SI), Fig. SI-1)) the pseudo-cubic c-lattice constant of the film, c=0.3769 nm, is very close to that of the single crystal, c=0.3773 nm, we believe the epitaxy strain can be ruled out. The bifurcation between ZFC and FC curves denotes the magnetic irreversibility in the system due to the presence of AFM/FM competing interactions among magnetic domains which is typical for the so called cluster glass behaviour. It means that along with Co$^{2+}$/Mn$^{4+}$ FM super exchange the AFM interactions of Mn$^{4+}$/Mn$^{4+}$ and/or Co$^{2+}$/Co$^{2+}$ type have to be present, although no additional features were observed at low temperatures due to the domain-wall de-pinning process similar to that observed in a single crystal[22]. The anomalous increase in FC curve at 18 K is due to the ordering of spins of Tb$^{3+}$ ions, which are likely FM-coupled to Co$^{2+}$/Mn$^{4+}$ sites.

The temperature dependence of reciprocal susceptibility, 1/χ(T), shown in Fig. 2(a), deviates from the linear Curie-Weiss behaviour for T<190 K and gradually decreases towards T$_C$. This indicates the appearance of short range ferromagnetic correlations among the spins in the overall paramagnetic background. The important role of this short range ordering will be discussed below by addressing magneto-dielectric coupling in the system. In Fig. 2(b) the field dependence of the isothermal magnetization, M(H), measured at 5 K is shown. The evaluated magnetization close to saturation magnetization, M$_S$~5.5 $\mu_B$/f.u., is significantly smaller than the theoretical value (6 $\mu_B$/f.u.) for a fully B-site ordered Co$^{2+}$/Mn$^{4+}$ system. Moreover, the Tb$^{+3}$ ions have even a higher moment of 9.72 $\mu_B$/Tb$^{3+}$ due to the spin-orbit coupling and, considering their additive contribution, the magnetic moment of the TCMO system should be very even much larger. Recently, the TCMO single crystal has been shown to have a strong magnetic anisotropy of Tb$^{3+}$ ions, which prefer to order along the c-axis yielding a large value of M$_S$~9.73 $\mu_B$/f.u.[22]. As for our TCMO/Nb:STO(001) film the c-axis according to XRD (Fig. SI-1) stays perpendicular to the film plane, the observed value of $M_S$~5.5 $\mu_B$/f.u. seems to be quite small. A very small value of remnant $M_r$=1.1 $\mu_B$/f.u and a large coercive field of $H_C$=0.35 T as compared to that of TCMO single crystal confirms a significant amount of $Mn^{3+}$ disorder along with $Co^{2+}$-O- $Co^{2+}$ and $Mn^{4+}$-O-$Mn^{4+}$ interactions, contributing to AFM phase boundaries and developing FM/AFM competitive interactions[12]. The presence of $Mn^{3+}$/ $Co^{3+}$ ions in the film has also been further confirmed by EELS mapping in HRTEM (Fig. SI-2). These disorder-induced short-range FM interactions are responsible for the deviation of 1/χ(T) from the Curie-Weiss law above $T_C$: the magnetic moments become coupled with electrical dipoles, originating from the same ions, as will be discussed in the following sections.

### B. Dielectric analysis

In Fig. 3(a) we present the temperature dependence of the real part of dielectric constant, $\epsilon'(T)$, measured for different frequencies, showing a broad maxima of $\epsilon'(T)$ which is of same order as observed in the LCMO film[14]. The temperature of the maximum, $T_m$, depends on the frequency and shifts to higher temperatures with increasing frequency. This kind of glassy behaviour is very new for the $A_2CoMnO_6$ double perovskites and probably indicates a ferroelectric relaxor behaviour, which is not associated with any structural transition in the system. With further lowering the temperature, the $\epsilon'(T)$ increases again possibly due to the electronic contribution from the conducting Nb:STO substrate and the substrate/film interface.

The observed ferroelectric relaxor behaviour can be fitted with the Curie-Weiss law, $\epsilon' = C/(T - \Theta)$ in the paraelectric region[23] above $T_m$ as shown in Fig. 3(b). The fitting parameters are presented in SI (see Tab. SI-1). At low frequencies the data were fitted well with the Curie constant, $C=8 \times 10^3$, and the Curie temperature, $\Theta$~150 K, respectively. With increasing frequency, the data start to deviate from the Curie-Weiss law. As for high frequencies the transition temperatures shift to higher side, the fitting range is not far away from the transition temperature. Similar to the magnetic case, the short range correlations among electric dipoles emerge, being frequency dependent and causing deviation from an ideal ferroelectric behaviour. The plausible reason is the formation of small Polar Nano Regions (PNR) having different response for sufficiently high frequency. For an ideal ferroelectric case the Curie-Weiss fit should not depend on the applied frequency.

In Fig. 3(c) one can see a broad relaxation peak ($T_m'$) in the temperature dependence of the dielectric loss, $\epsilon''(T)$, which also shows a frequency dispersion. The observed relaxation in both real, $\epsilon'(T)$, and imaginary part, $\epsilon''(T)$, of dielectric constant could be understood in the framework of a dipolar glass model[24]. It considers small dipolar regions, induced in the system by the B-site disorder, the dipole moment of which fluctuates/vibrates thermally. Analogically to the spin glass the dipoles are expected to be frozen at low temperatures. The freezing temperature ($T_f$) is finite if the interaction among the dipoles are strong enough. The dynamics of the PNR is described by the so called Vogel-Fulcher (VF) formalism $f^{-1} = \tau_0 exp[U_a/K_B(T_{max} - T_f)]$, where f is the frequency of an applied electric field, the $T_{max}$ is the peak maxima[25–28]. If the electrostatic interaction among the dipoles is not strong enough to freeze them cooperatively, then dipoles can vibrate with external ac electric field at any finite temperature and can show a thermally activated Arrhenius behaviour with $T_f \rightarrow 0K$.

Both VF and the Arrhenius law $f^{-1} = \tau_0 exp[U_a/K_B(T_{max})]$ along with a power law were tried to fit the data. The best fit was obtained as Arrhenius behaviour, shown in Fig. 3(d) for both $\epsilon'$ and $\epsilon''$. The activation energy, calculated from the fit, is $U_a$=1.9 eV which is very large for a typical relaxor ferroelectric[24]. This anomalously large value of $U_a$ and the deviation from a typical VF law have been further investigated from the asymmetry in the relaxor peak in $\epsilon'(T)$. According to the diffuse phase transition model, the temperature dependent dielectric permittivity as well as the size distribution of PNR's can be described by a Gaussian distribution function $\frac{1}{\sqrt{2\pi\sigma^2}}\exp[-\frac{(T-T_m)^2}{2\sigma^2}]$ [24]. Our experimental data can be fitted well by this distribution function and the overall behaviour of the temperature dependent relaxor peak in $\epsilon'$ was found to be a superposition of three different maxima, denoted as $T_m^1$, $T_m^2$ and $T_m^3$, shown in Fig. 4(a) for f=10 kHz. The data for all frequencies are fitted and these three frequency dependent temperatures are presented in SI, Tab. SI-2. Again VF, activation law and power law were tried to fit these three temperatures and we found the Arrhenius behaviour provides the best fits as shown in Fig. 4 (b), (c) and (d). The calculated activation energies, 0.22, 0.30 and 0.32 eV, are now physically reasonable. From the deconvolution of the $\epsilon'$ peaks we obtained three types of PNRs, which differ in their microscopic origin and have distinguishable size distributions. These three distinct classes of dipoles can be related to their microscopic origin, taking into account that electric dipoles in TCMO may originate from the oxygen bonds with $Co^{2+}$ and $Mn^{4+}$ ions in the $CoO_6$ and $MnO_6$ octahedra, which are getting distorted in an applied electric field. The third contribution could be caused by the disorder-induced presence of the $Mn^{3+}$ as well as of $Co^{3+}$ ions. All these PNRs started to interact with lowering temperature. Due to the difference in their distributions as well as in the activation energies they respond differently with external frequency and with temperature. The overall macroscopic response is, hence, showing a ferroelectric relaxation.

In order to inspect the detail interactions among these dipoles dielectric loss has been studied in the frequency domain for different temperatures. For the used frequency range, f=20 Hz-2MHz, the two main mechanisms can be responsible for the dipole relaxation: (1) the Maxwell-Wagner mechanism because of the local charge accumulation and grain boundaries[29] and (2) the Debye relaxation which is the dipolar contribution from the hopping of charge carriers among asymmetric sites ($Mn^{4+}$, $Co^{2+}$ and $Mn^{3+}$)[30]. Fig. 5(a) shows the dielectric relaxation over the frequency range for the temperatures in the relaxor transition regime. At very low frequency regime the data shows a Maxwell-Wagner behaviour which looks like a monotonic linear decrease of $\epsilon''$ with frequency in logarithmic scale. With increasing frequency the data start to deviate from the Maxwell-Wagner behaviour and display two distinct Debye relaxation peaks, denoted as $\tau_1$ and $\tau_2$. These two relaxation behaviours could be recognized as β($\tau_1$)- and α($\tau_2$)-like processes in a glassy system[31]. The α relaxation is a principal relaxation process in this dipolar glassy system originating from the major charge transfer between the $Co^{2+}$-$Mn^{4+}$ sites. A secondary or β process primarily develops from the localized defects or minor sites such as $Mn^{3+}$. Interestingly we observed that this two processes merge around T*~190 K giving a single relaxation peak as presented in Fig. 5(b). The deviation from the Curie-Weiss law in magnetic behaviour at the same temperature points out onto a correlation between spin and polar order parameter that will be analysed in detail in the next section.

### C. Magneto-Dielectric Coupling

Magneto-dielectric (MD) analysis has been done by dielectric measurements in an applied magnetic field, B=8T, at f=100 kHz. Here we can observe a characteristic change in the dielectric constant close to the relaxor transition as shown in Fig. 6 (a) The temperature dependence of the MD coupling constant $\epsilon_{MD} = \frac{\epsilon'(8T)-\epsilon'(0T)}{\epsilon'(0T)} \times 100$ along with the derivative of dielectric constant ($\frac{\delta\epsilon'}{\delta T}$) in Fig. 6 (b) shows that $|\epsilon_{MD}|$ increases by cooling down the system and takes the highest value of 4 % at the relaxor transition. After that it decreases again with cooling while dipoles are started freezing. Interestingly, this occurs in the same region where a short range spin ordering develops within the magnetic domain (see the inset in Fig. 6 (a)). The absence of any magnetoresistance (see Fig. SI-3 in SI) confirms that the MD coupling is intrinsic to the dipoles present in the material. As indicated previously, mostly the Debye processes[30], resulting from of charge transfers among dipoles, contribute to dielectric constant at high frequency. Dielectric constant increases up to the relaxor transition mainly from the contribution of activated dipoles due to the charge transfer from $Mn^{3+}$ to other sites. Bellow the relaxor transition $Mn^{4+}$ and $Co^{2+}$ start interacting magnetically with the other disorder sites as seen from the short range correlation developed bellow 190 K and dielectric relaxation peak splits at the same temperature indicating a charge transfer between them. As the charge transfer is coupled to and depends on the spin arrangement, the short range spin-spin interaction tries to restrict the charge hopping and we can see the 2$^{nd}$ peak is not that much pronounced as the first one. With applied strong magnetic field this process further interrupted and dielectric constant further reduce, causing a 4% MD coupling at the same temperature.

## IV. CONCLUSION

In summary we have grown monoclinic phase of $Tb_2CoMnO_6$ double perovskite thin film on Nb:$SrTiO_3$ (100) by using MAD technique. TEM/EELS mapping shows the presence and distribution of both $Co^{2+}$ as well as $Co^{3+}$ ions in the film, evidencing a partial B-site disorder, further confirmed by the observed reduction of the saturation magnetization at low temperatures. The ferromagnetic $T_C$=110 K was slightly higher as compared to the bulk value due to an in plane tensile strain. The presence of a short range ordering of spins at T~190 K>>$T_C$ was established that causes the deviation from standard Curie-Weiss Law far above the ferromagnetic $T_C$. Two different dielectric relaxation peaks (β and α) have been observed that merge at the same temperature of short range spin correlation. Moreover, we observed an unexpected high temperature dipolar relaxor-glass-like transition, at which a coupling to short range magnetic correlations results in a 4% magneto-dielectric coupling.

**ACKNOWLEDGEMENTS**

R.M. acknowledges financial support from Erasmus Plus programme, European Union, Georg-August-Universität Göttingen and IISER Pune. R.M. and M.C. are thankful to Mr. Satish Yadav for helping in Dielectric measurement. V.R. and V.M. acknowledge financial support from Deutsche Forschungsgemeinschaft (DFG) via SFB 1073 [TP A02, TP Z02] as well as via DFG Project MO-2254-4.

**REFERENCES**

[1] Y. Tokura, S. Seki, and N. Nagaosa, Reports Prog. Phys. **77**, 076501 (2014).

[2] N.A. Spaldin and R. Ramesh, Nat. Mater. **18**, 203 (2019).

[3] M. Bibes and A. Barthélémy, Nat. Mater. **7**, 425 (2008).

[4] M. Coll, et al, Appl. Surf. Sci. **482**, 1 (2019).

[5] S. Gariglio, A.D. Caviglia, J.M. Triscone, and M. Gabay, Reports Prog. Phys. **82**, 012501 (2019).

[6] M. Bibes and A. Barthélémy, IEEE Trans. Electron Devices **54**, 1003 (2007).

[7] L.W. Martin, Y.H. Chu, and R. Ramesh, Mater. Sci. Eng. R Reports **68**, 89 (2010).

[8] M.K. Kim, J.Y. Moon, H.Y. Choi, S.H. Oh, N. Lee, and Y.J. Choi, J. Phys. Condens. Matter **27**, 426002 (2015).

[9] Y. Shimakawa, M. Azuma, and N. Ichikawa, Materials (Basel). **4**, 153 (2010).

[10] M. Zhu, Y. Lin, E.W.C. Lo, Q. Wang, Z. Zhao, and W. Xie, Appl. Phys. Lett. **100**, 062406 (2012).

[11] C. Meyer, V. Roddatis, P. Ksoll, B. Damaschke, and V. Moshnyaga, Phys. Rev. B **98**, 134433 (2018).

[12] R.I. Dass and J.B. Goodenough, Phys. Rev. B - Condens. Matter Mater. Phys. **67**, 014401 (2003).

[13] S. Baidya and T. Saha-Dasgupta, Phys. Rev. B - Condens. Matter Mater. Phys. **84**, 035131 (2011).

[14] M.P. Singh, K.D. Truong, and P. Fournier, Appl. Phys. Lett. **91**, 042504 (2007).

[15] D. Choudhury, P. Mandal, R. Mathieu, A. Hazarika, S. Rajan, A. Sundaresan, U. V. Waghmare, R. Knut, O. Karis, P. Nordblad, and D.D. Sarma, Phys. Rev. Lett. **108**, 127201 (2012).

[16] C. Xie, L. Shi, J. Zhao, S. Zhou, Y. Li, and X. Yuan, J. Appl. Phys. **120**, 0 (2016).

[17] J. Blasco, J. García, G. Subías, J. Stankiewicz, S. Lafuerza, J.A. Rodríguez-Velamazán, C. Ritter, and J.L. García-Muñoz, J. Phys. Condens. Matter **26**, 386001 (2014).

[18] J. Blasco, J. García, G. Subías, J. Stankiewicz, J.A. Rodríguez-Velamazán, C. Ritter, J.L. García-Muñoz, and F. Fauth, Phys. Rev. B **93**, 214401 (2016).

[19] V. V. Shvartsman, S. Bedanta, P. Borisov, W. Kleemann, A. Tkach, and P.M. Vilarinho, Phys. Rev. Lett. **101**, 165704 (2008).

[20] J.H. Lee, P. Murugavel, H. Ryu, D. Lee, J.Y. Jo, J.W. Kim, H.J. Kim, K.H. Kim, Y. Jo, M.H. Jung, Y.H. Oh, Y.W. Kim, J.G. Yoon, J.S. Chung, and T.W. Noh, Adv. Mater. **18**, 3125 (2006).

[21] M. Jungbauer, S. Hühn, R. Egoavil, H. Tan, J. Verbeeck, G. Van Tendeloo, and V. Moshnyaga, Appl. Phys. Lett. **105**, 251603 (2014).

[22] J.Y. Moon, M.K. Kim, D.G. Oh, J.H. Kim, H.J. Shin, Y.J. Choi, and N. Lee, Phys. Rev. B **98**, 174424 (2018).

[23] M. Tyunina and J. Levoska, Phys. Rev. B - Condens. Matter Mater. Phys. **70**, 132105 (2004).

[24] C.W. Ahn, C.H. Hong, B.Y. Choi, H.P. Kim, H.S. Han, Y. Hwang, W. Jo, K. Wang, J.F. Li, J.S. Lee, and I.W. Kim, J. Korean Phys. Soc. **68**, 1481 (2016).

[25] A.A. Bokov and Z.G. Ye, Front. Ferroelectr. A Spec. Issue J. Mater. Sci. **1**, 31 (2007).

[26] D. Viehland, S.J. Jang, L.E. Cross, and M. Wuttig, J. Appl. Phys. **68**, 2916 (1990).

[27] A.E. Glazounov and A.K. Tagantsev, Appl. Phys. Lett. **73**, 856 (1998).

[28] L. Eric Cross, Ferroelectrics **76**, 241 (1987).

[29] A. R. von Hippel, Dielectrics and Waves (MIT Press, Cambridge, MA, 1966).

[30] K. C. Kao, Dielectric Phenomena in Solids (Elsevier Academic Press, London, 2004).

[31] F. Kremer and A. Schönhals, *Broadband Dielectric Spectroscopy* (2003)

**FIGURE CAPTIONS**

**Fig. 1.** Zero- (green, ZFC) and field-cooled (red, FC) magnetic susceptibility ($\chi(T)$) as a function of temperature (T) in the TCMO film.

**Fig. 2.** (a) $1/\chi(T)$ vs T plot indicating the temperature for short range spin correlation $T_S$ with the deviation from Curie-Weiss behaviour (inset). (b) Isothermal magnetization with applied external magnetic field at 5K.

**Fig. 3.** (a) Dielectric Constant ($\epsilon'(T)$) vs Temperature with different applied frequencies(1-100 KHz). (b) Curie-Weiss fit for the $\epsilon'(T)$ above the relaxor glass transition fro different frequencies. (c) Temperature dependence of the Dielectric Loss part $\epsilon''(T)$ with different applied frequencies showing relaxation peaks. (d) Arrhenius fit for the relaxation time vs peak temperature for the Dielectric Constant $\epsilon'(T)$ and Dielectric Loss $\epsilon''(T)$ (inset).

**Fig. 4.** (a) Gaussian fit for the temperature dependent Dielectric Constant $\epsilon'(T)$ and deconvolution of three peaks signifying the presence of three different type of PNR s. (b),(c) and (d) Arrhenius fit for three relaxation times from these three different peaks $T_m^1$, $T_m^2$ and $T_m^3$.

**Fig. 5.** (a) Frequency dependence of the Dielectric Loss part $\epsilon''(T)$ at different temperatures in the frequency range of f=20 Hz-2MHz. (b) Temperature dependence of the two deferent Debye relaxation processes $\tau_1$ and $\tau_2$.

**Fig. 6.** (a) Temperature dependence of the Dielectric Constant $\epsilon'(T)$ at 100KHz without and with applied external magnetic field of 8T. (b) Temperature dependence of derivative of Dielectric Constant ($\frac{\delta\epsilon'}{\delta T}$) along with the magneto-dielectric coupling constant $\epsilon_{MD}$ (%) showing a coupling between relaxor dipolar glassy transition with short range spin correlation resulting maximum 4% magneto-dielectric coupling.

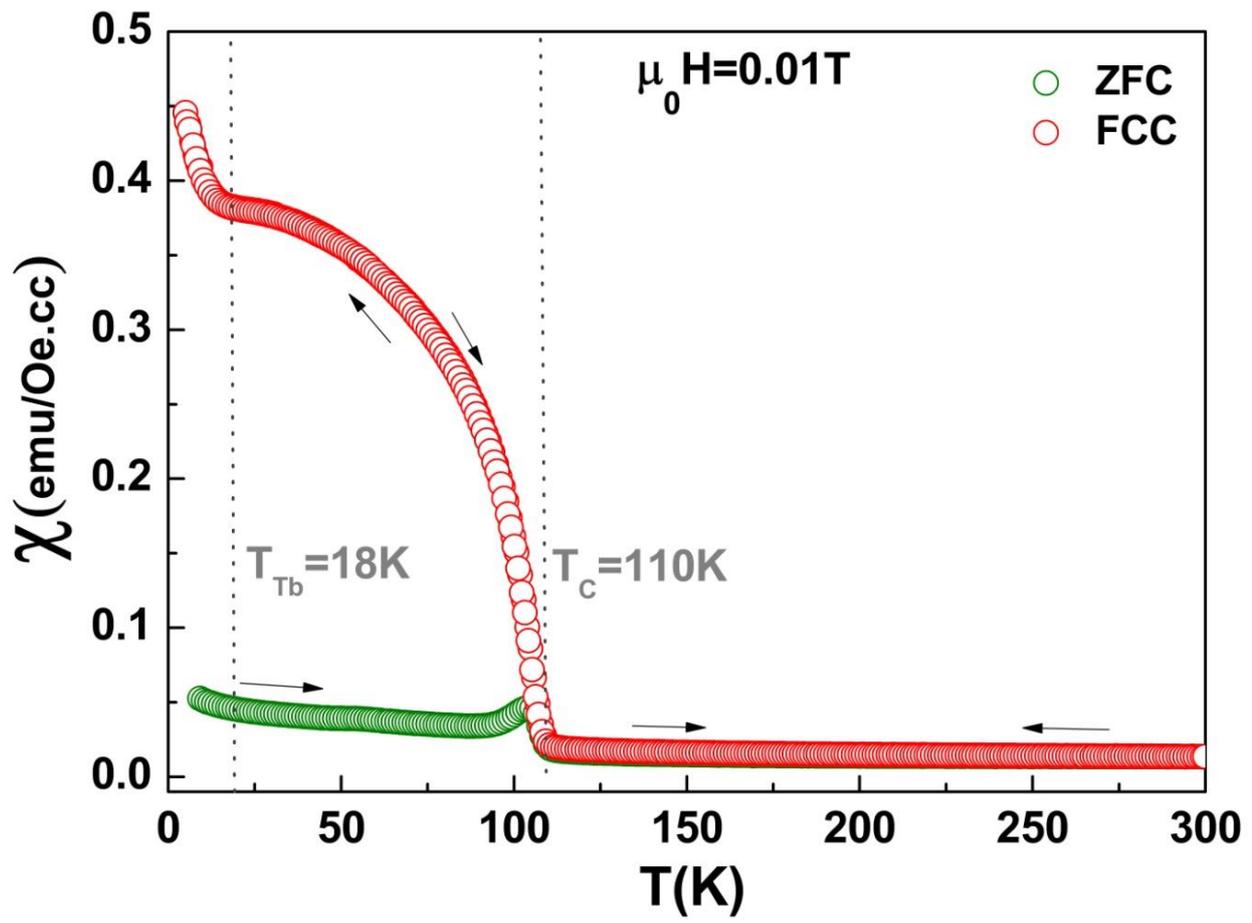

**Figure 1**

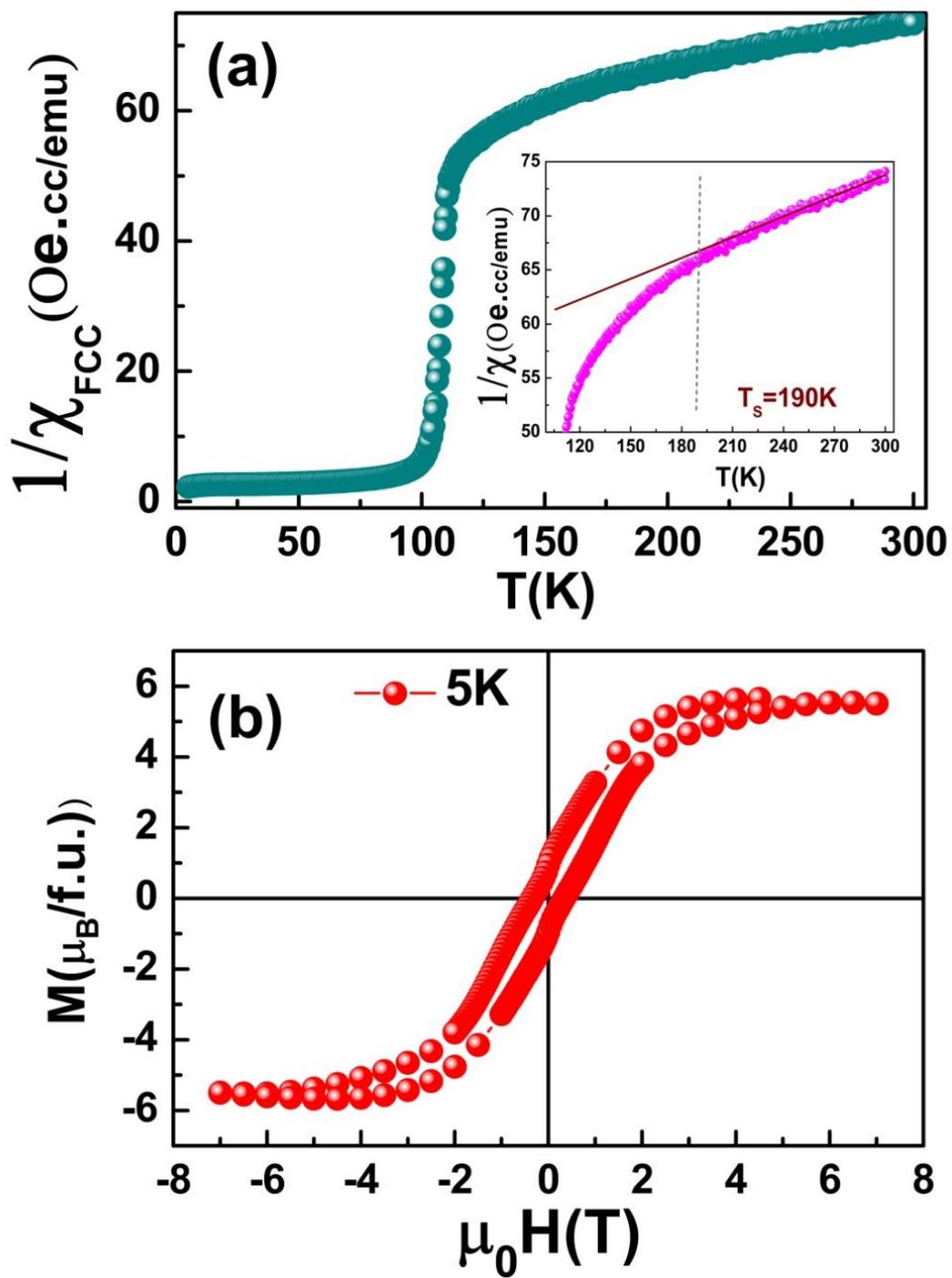

**Figure 2**

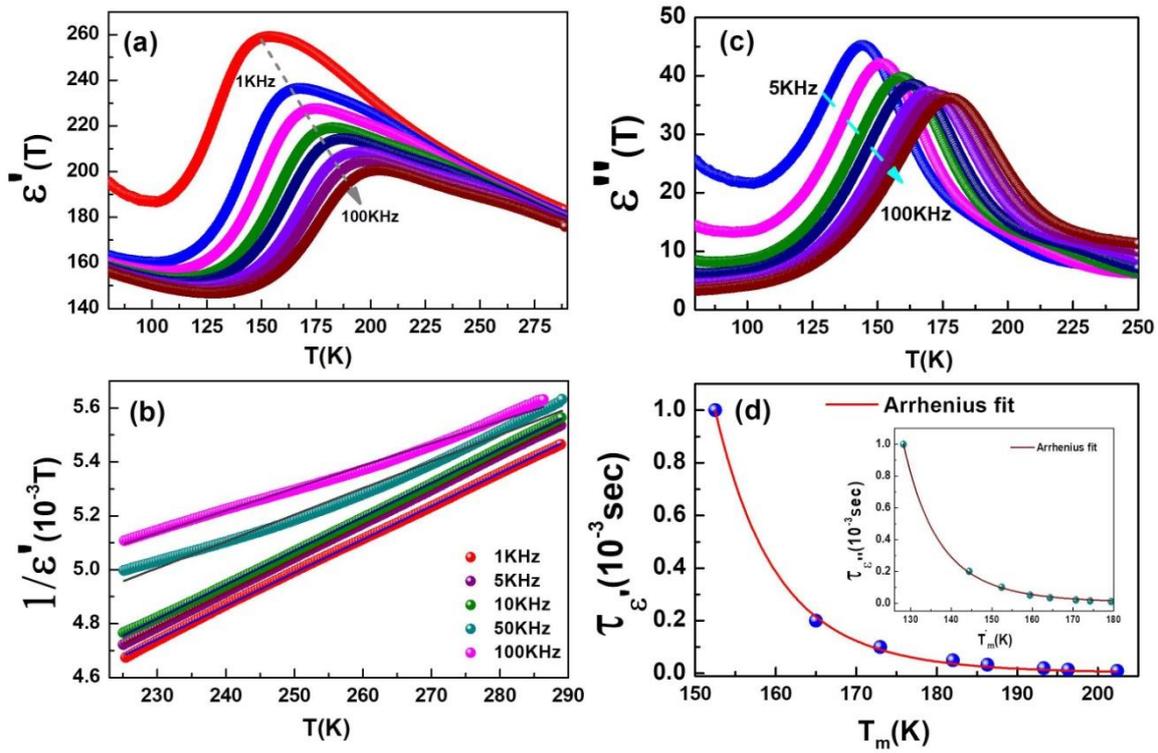

Figure 3

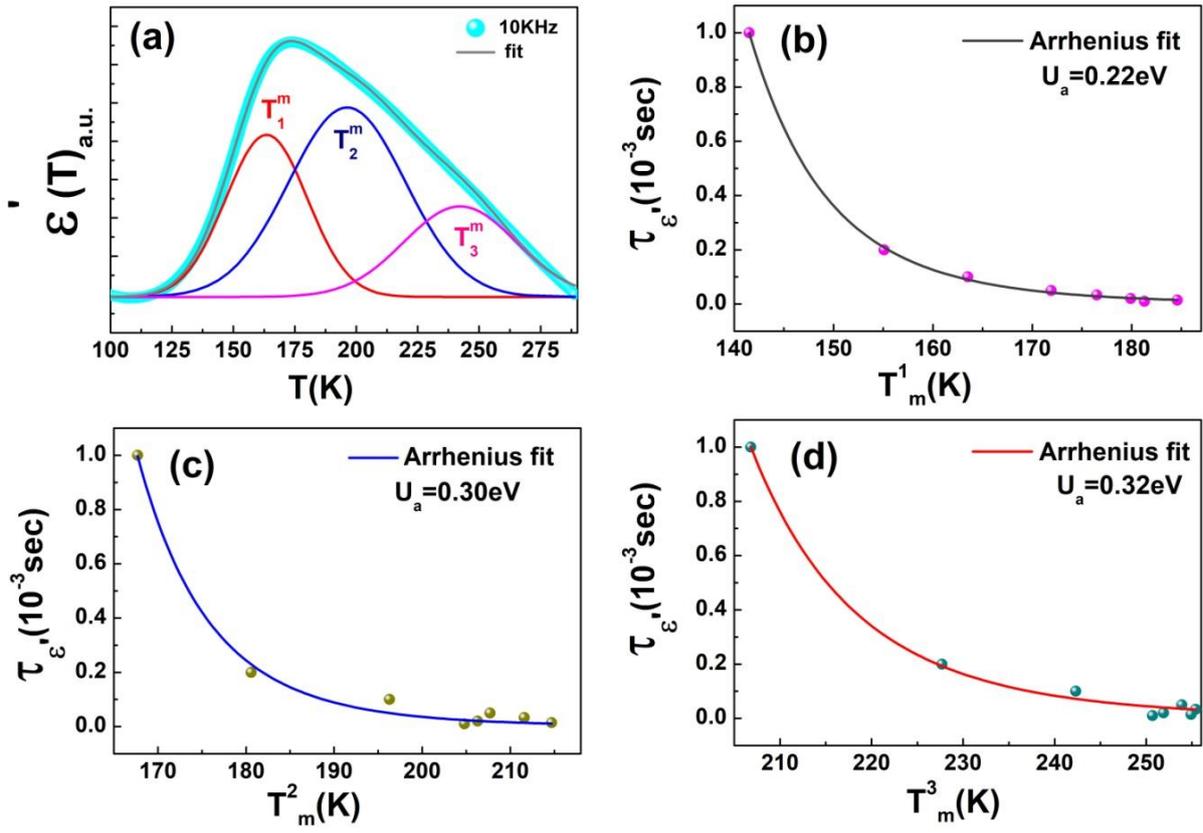

**Figure 4**

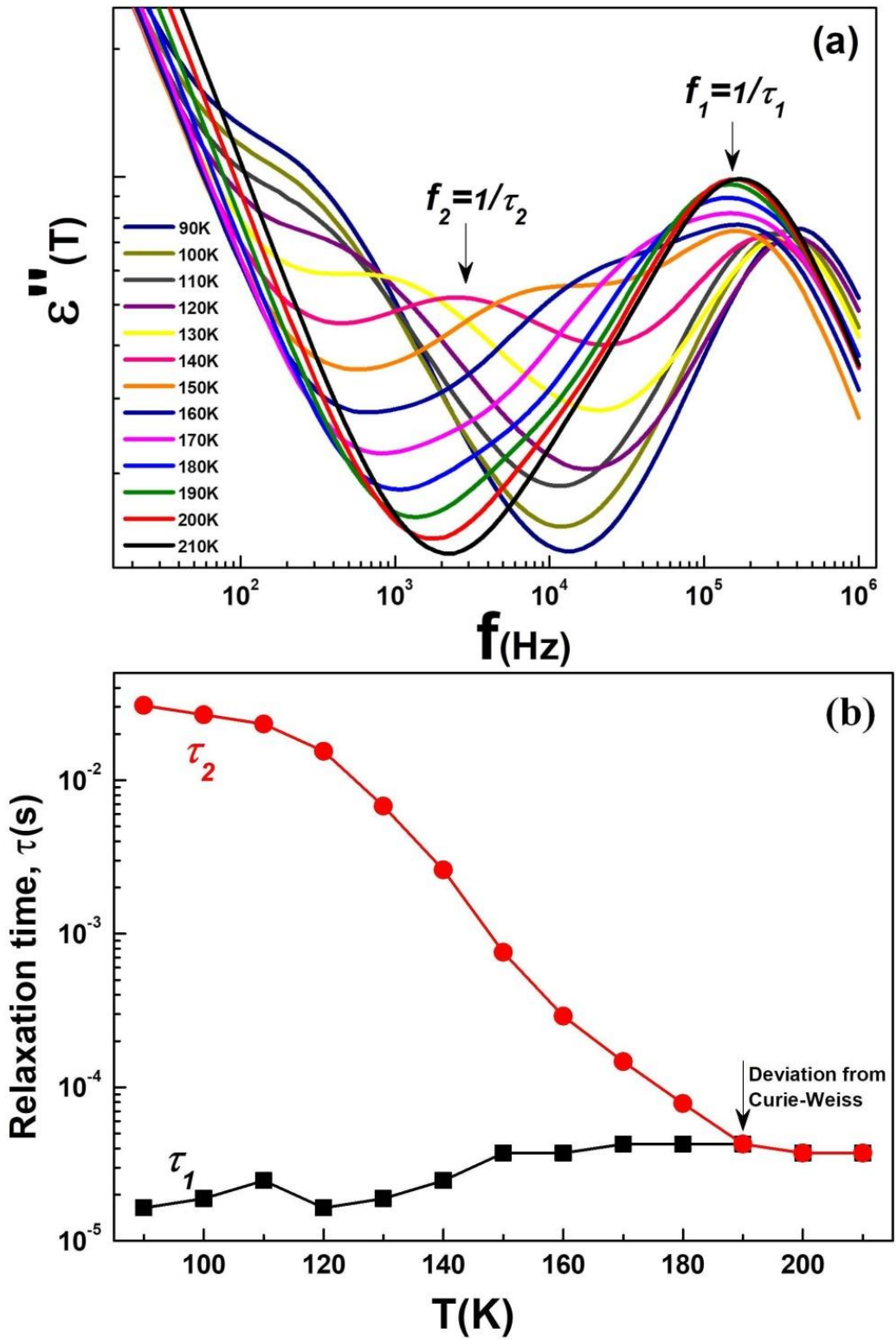

**Figure 5**

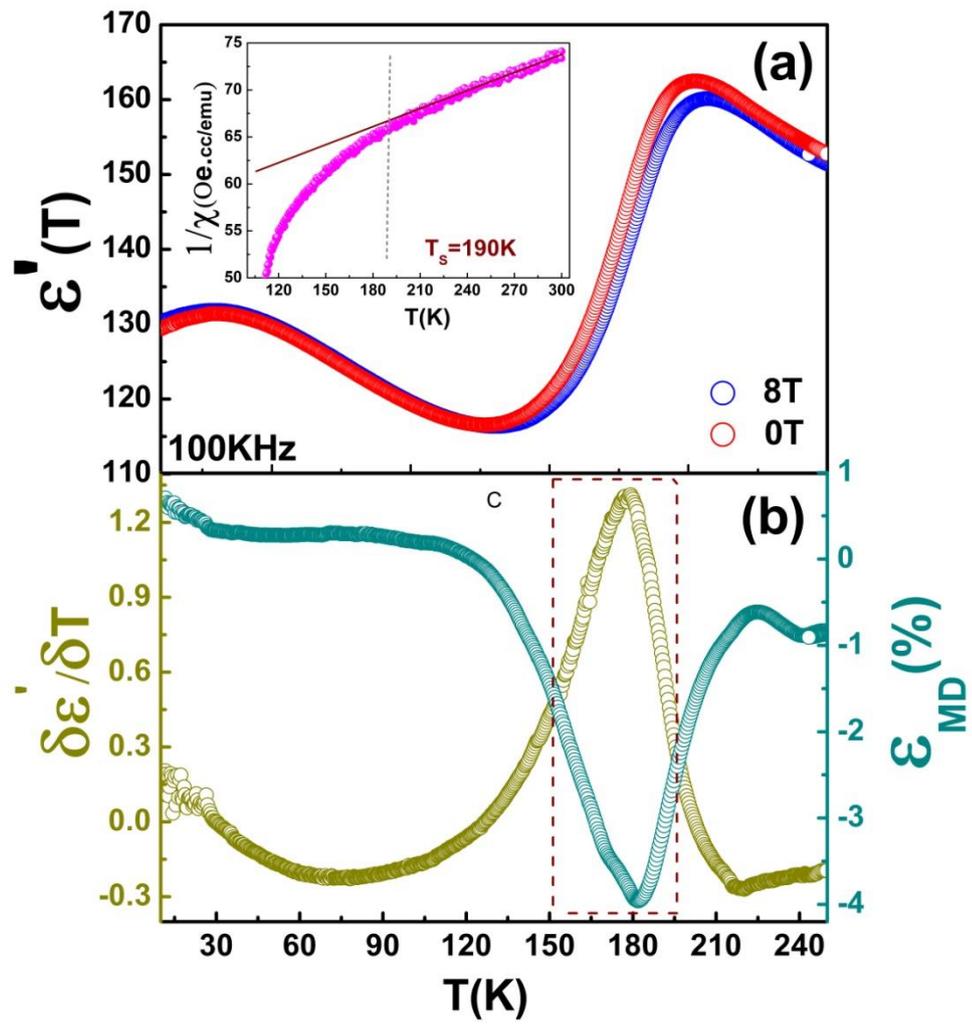

**Figure 6**

# Supplementary Information

## Magneto-dielectric Effect in relaxor dipolar glassy $Tb_2CoMnO_6$ Film

R. Mandal[1,2*], M. Chandra[3], V. Roddatis[4], M. Tripathi[3], R. Rawat[3], R.J. Choudhary[3] and V. Moshnyaga[2*]

[1]*Department of Physics, Indian Institute of Science Education and Research, Pune 411008, India*

[2]*Erstes Physikalisches Institut, Georg-August-Universität Göttingen, Friedrich-Hund-Platz 1, 37077 Göttingen, Germany*

[3] *UGC-DAE Consortium for Scientific Research, Indore Centre, University Campus, Khandwa Road, Indore 452017, India*

[4]*Institut für Materialphysik, Georg-August-Universität Göttingen, Friedrich-Hund-Platz 1, 37077 Göttingen, Germany*

*E-mail; vmosnea@gwdg.de ; rajesh.mandal@students.iiserpune.ac.in*

**Table SI-1.** Curie-Weiss Fit parameters

| Frequency(KHz) | Curie Constant | $\Theta(K)$ | $T_m(K)$ |
| --- | --- | --- | --- |
| 1 | $8.1 \times 10^3$ | 154.8 | 152.6 |
| 5 | $7.9 \times 10^3$ | 146.4 | 166.6 |
| 10 | $8.0 \times 10^3$ | 153.4 | 172.5 |
| 50 | $10 \times 10^3$ | 277.2 | 192.6 |
| 100 | $12 \times 10^3$ | 388.7 | 202.5 |

**Table SI-2.** Peak temperatures from fitting of temperature dependent dielectric maxima($\epsilon'$)

| Frequency(KHz) | $T_m^1$(K) | $T_m^2$(K) | $T_m^3$(K) |
|---|---|---|---|
| 1 | 141.5 | 167.7 | 206.8 |
| 5 | 155.1 | 180.6 | 227.7 |
| 10 | 163.5 | 196.3 | 242.3 |
| 20 | 171.9 | 207.7 | 253.9 |
| 30 | 176.5 | 211.6 | 255.4 |
| 50 | 179.9 | 206.3 | 251.9 |
| 70 | 184.6 | 214.7 | 254.9 |
| 100 | 181.3 | 204.8 | 250.7 |

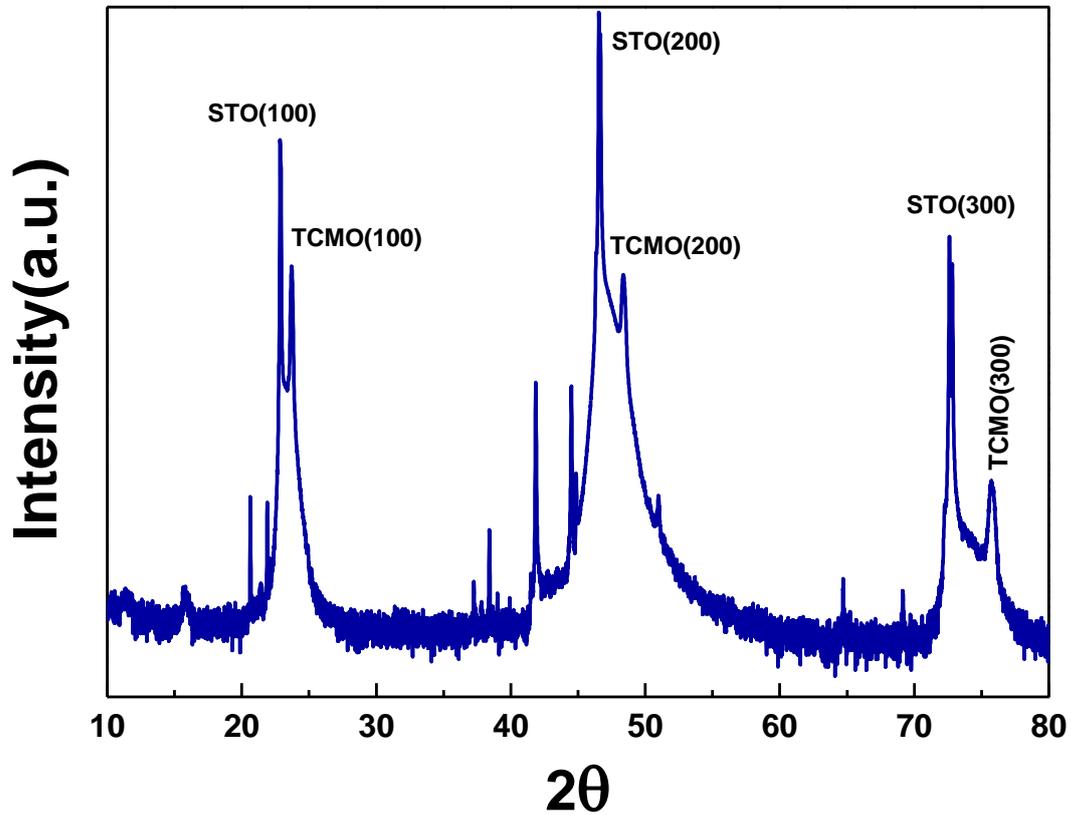

**Figure SI-1.** X-ray diffraction Θ-2Θ pattern for $Tb_2CoMnO_6$/Nb:$SrTiO_3$ (100). The evaluated out of plane pseudocubic lattice parameter of TCMO, $C_{film}$=0.3769 nm, is very close to that of bulk single crystal, C=0.37735 nm (ref. 22).

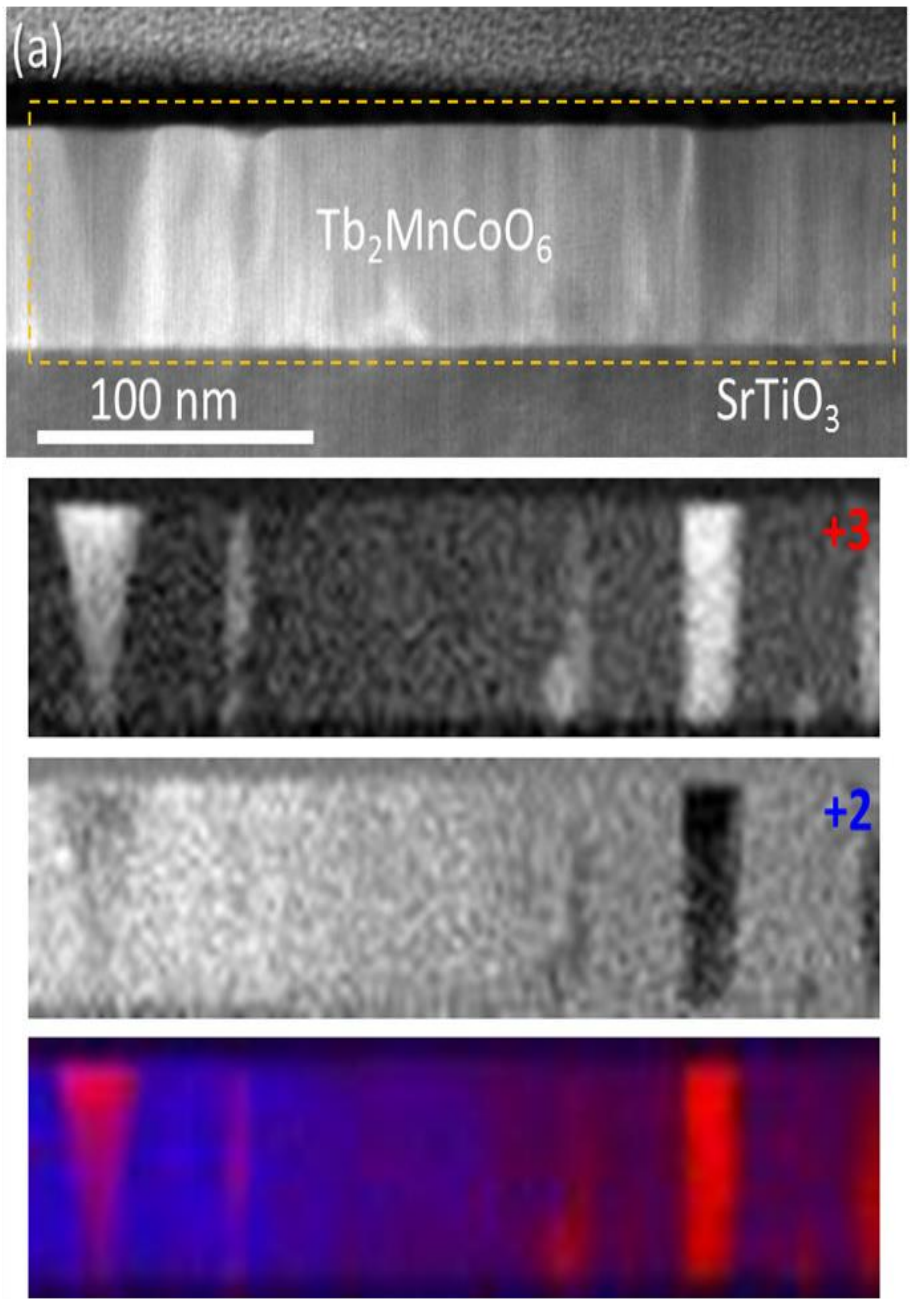

**Figure SI-2**.TEM/EELS mapping on TCMO film. The distribution of Co2+ and Co3+ within a region in the film. One can see that some regions have excess of Co ions (white regions on the +3 panel) and these regions have Co3+ ions. The rest is mostly containing Co2+, i.e. the ordered TMCO. Mn distribution is overall homogeneous (no excess of Mn ions). However, it is not easy to make a distribution of Mn3+ and Mn4+ because they do not differ strongly from each other.

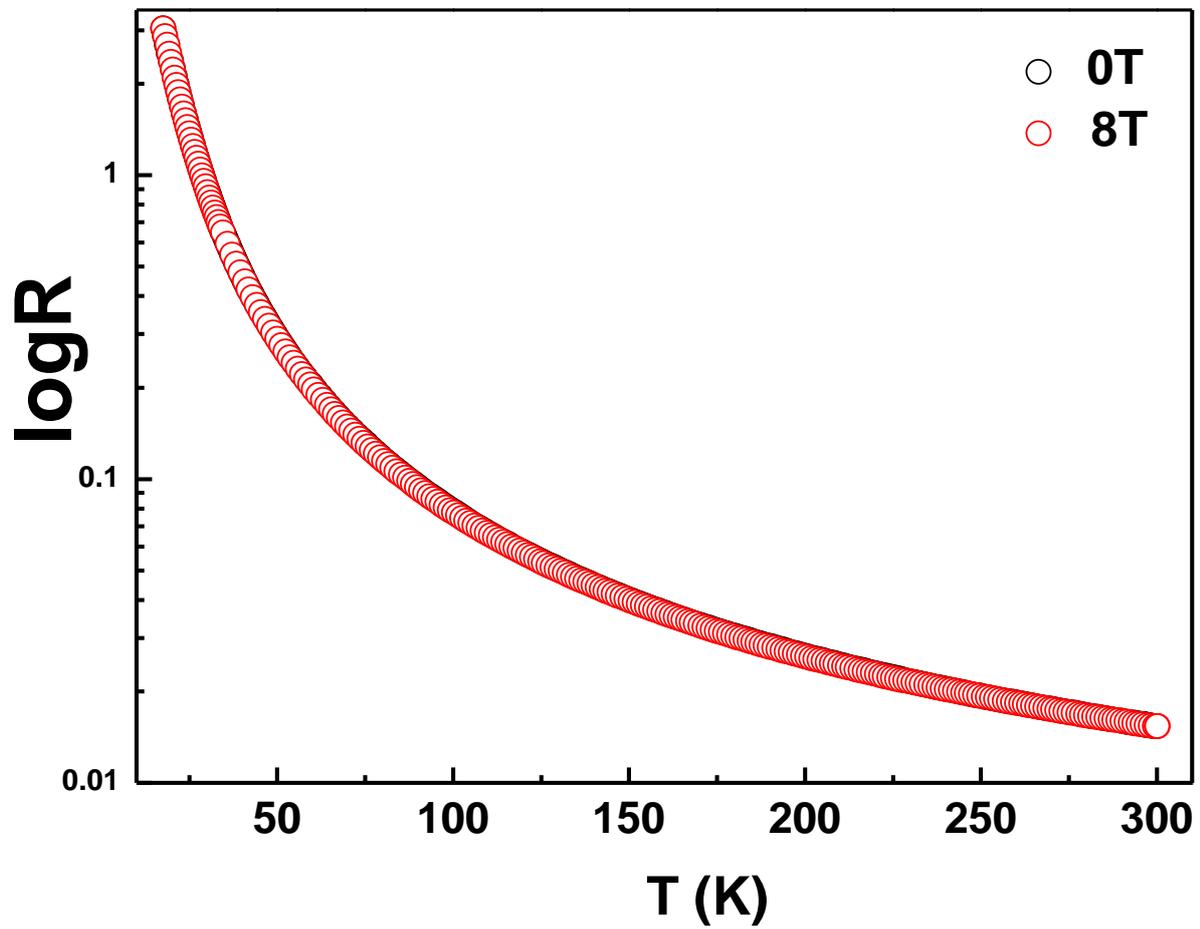

**Figure SI-3.** Temperature dependences of DC resistance, measured in external magnetic field B=0 and B=8T, evidence the absence of any magnetoresistance.